\documentclass[%
reprint,
amsmath,amssymb,
aps,
showkeys
]{revtex4-2}
\usepackage[braket, qm]{qcircuit}
\usepackage{graphicx}
\usepackage{dcolumn}
\usepackage{bm}
\usepackage{hyperref}
\hypersetup{colorlinks,
allcolors=blue}
\usepackage{float}
\usepackage[braket, qm]{qcircuit}
\usepackage{graphicx}
\usepackage{multirow}
\usepackage{array}
\usepackage{tikz}
\usepackage{soul}
\begin{document}
\title{A Hardware-Efficient M{\o}lmer-S{\o}rensen Gate for Superconducting Quantum Computers}
\author{M. AbuGhanem{$^{1}$}}
\address{$^{1}$ Faculty of Science, Ain Shams University, Cairo, $11566$, Egypt} 
\email{gaa1nem@gmail.com; \\ 
mghanem@sci.asu.edu.eg}
\date{\today}
\begin{abstract}
The Mølmer-Sørensen gate, a cornerstone entangling operation in trapped-ion systems, represents a promising alternative to standard entangling gates in superconducting quantum architectures. However, its performance on superconducting hardware has remained unverified. In this work, we present a hardware-efficient implementation of the Mølmer-Sørensen gate and characterize its performance using quantum process tomography (QPT) on IBM Quantum's superconducting processors. Our implementation achieves a process fidelity of 92.47\% on the real quantum hardware, a performance competitive with the 93.02\% fidelity of the device's native controlled-NOT (CX) gate. Furthermore, for the $\ket{00}$ input state, the gate prepares the target Bell state with $94.2\%$ success probability, confirming its correct logical operation. These results demonstrate that non-native entangling gates can be optimized to perform on par with hardware-native operations. This work expands the effective gate set for algorithm design on fixed-architecture processors and provides a critical benchmark for cross-platform gate evaluation, underscoring the role of hardware-aware compilation in advancing noisy intermediate-scale quantum (NISQ) computing.
\end{abstract}

\keywords{
Quantum Gate Benchmarking, 
Mølmer-Sørensen Gate, 
Quantum Process Tomography, 
Superconducting Qubits, 
Quantum Compilation, 
NISQ Era
IBM's Quantum computers  \\
PACS: 
$03.67.Lx$, 
$03.67.-a$,  
$03.67.-a$, 
$03.65.Ca$ \\
}

\maketitle

\section{Introduction}

Quantum computation represents a paradigm shift in information processing~\citep{Light}, promising to solve classically intractable problems in areas such as pharmaceutical~\citep{Drugdesign}, material science~\citep{materials-science}, and cryptography~\citep{Shor,factor2048bitRSA}, among others~\citep{IBMQuantum}. This potential stems from the unique principles of quantum mechanics---superposition, interference, and entanglement---which allow quantum algorithms to explore computational paths in parallel~\citep{Theprinciples,AbuGMSc19}. The realization of this potential, however, critically depends on the precise control of quantum bits (qubits) and the high-fidelity execution of quantum logic gates~\citep{DiVincenzo}. Within this framework, the quality of two-qubit entangling gates often serves as the primary bottleneck for overall computational performance, as they are essential for creating the entangled states that power quantum advantage~\citep{Preskill2012}.

The computational advantage of quantum information processing~\citep{NISQ24} derives fundamentally from entanglement~\citep{Preskill2012}, a resource enabled by high-fidelity two-qubit operations~\citep{Kim23,Willo25}. While single-qubit gates provide the foundational rotations of the quantum state space, two-qubit gates create the correlations necessary for quantum parallelism and interference~\citep{Barenco1995}. In the circuit model of quantum computation, these entangling operations, combined with single-qubit rotations, form a universal gate set capable of implementing any quantum algorithm~\citep{MikeIke}.

The landscape of two-qubit gates is characterized by a trade-off between expressivity and hardware compatibility. The CNOT gate serves as a canonical example, performing a conditional bit-flip operation that creates maximal entanglement from separable input states~\citep{Barenco1995}. However, various physical platforms naturally implement different entangling interactions, leading to the development of platform-specific gate sets. Superconducting quantum processors~\citep{OQCtransmon,SuperconductingQuantum} typically natively implement either the CNOT gate or its equivalent, the controlled-Z (CZ) gate, through controlled-phase interactions~\citep{DiCarlo2009}. Meanwhile, trapped-ion systems leverage their collective vibrational modes to realize different entangling paradigms~\citep{MS-1,MS-2}, among which the Mølmer-Sørensen (MS) gate stands as a particularly significant contribution~\citep{MS-1}.

The Mølmer-Sørensen gate was originally conceived to overcome experimental challenges in laser-driven trapped-ion quantum computation~\citep{MS-1,MS-2,MS-3}. Where previous approaches required precise ground state cooling and were sensitive to thermal motion~\citep{CiracZoller1995}, the Mølmer-Sørensen gate operates effectively outside the Lamb-Dicke regime and is resilient to motional heating~\citep{Roos2008}. This robustness stems from its geometric phase-based mechanism, which entangles ion qubits through their shared interaction with a common motional mode, driven by laser fields detuned from the sideband transitions~\citep{Leibfried2003, Monroe2021}.

As the field advances toward hardware-agnostic quantum programming, the efficient compilation of non-native gates across hardware platforms has become a critical task in quantum compiler optimization~\citep{EmuPlat,KetPlatform,AQUA}. The Mølmer-Sørensen gate presents a compelling case for such cross-platform translation due to its proven utility and theoretical robustness in trapped-ion systems~\citep{Akerman2015, QPT-MS}. However, the practical performance of compiled Mølmer-Sørensen gates on superconducting hardware has not been systematically benchmarked against native entangling operations.

In this work, we develop and characterize a hardware-efficient implementation of the Mølmer-Sørensen gate for superconducting quantum processors. We provide a comprehensive fidelity benchmark against native two-qubit gates, using quantum process tomography. Our results demonstrate that the compiled MS gate achieves a fidelity comparable to the native CX gate. This finding indicates that through careful circuit decomposition and optimization, non-native gates can compete with hardware-native operations, thereby expanding the effective gate set for quantum algorithm design on fixed-architecture superconducting processors~\citep{Chow2011}. Our work has significant implications for quantum compiler design and the effective utilization of diverse gate sets in the era of NISQ computing~\citep{NISQ18, VQA}.

The rest of this paper is organized as follows: Section~\ref{sec:Methods} details our methods, including the hardware-efficient implementation of the Mølmer-Sørensen gate (\ref{sec:Hardware-Efficient}), its validation via direct state measurements (\ref{sec:StateMeasurements}), and its comprehensive characterization using quantum process tomography (\ref{sec:Complete-Characterization}). Section~\ref{sec:Results} presents our results, analyzing the implementation's logical efficiency (\ref{sec:logical-efficiency}) and benchmarking its experimental fidelity against the native CX gate (\ref{sec:fidelity}). Section~\ref{sec:implications} discusses the broader implications of our findings. Section~\ref{sec:Analysis-Hardware} provides an analysis of the underlying quantum hardware performance, and Section~\ref{sec:conclusion} concludes the paper.

\section{Methods}
\label{sec:Methods}

\subsection{Hardware-Efficient Implementation}
\label{sec:Hardware-Efficient}

The diversification of quantum computing platforms has led to a variety of physical implementations, each with a distinct native gate set~\citep{SuperconductingQuantum, Neutral-atom-epj,IonQ-30qubits,PhotonicQuantumComputers,GoogleAI}. The Mølmer-Sørensen gate, originally developed for trapped-ion systems~\citep{MS-1,MS-2,MS-3}, is a maximally entangling operation that is locally equivalent to the CX gate. While the CX gate is the standard two-qubit operation on superconducting processors, the efficient compilation of non-native gates like the MS gate can enrich the available gate repertoire, potentially enabling more efficient quantum circuits and algorithm-specific optimizations.

Mathematically, the Mølmer-Sørensen gate implements the unitary transformation:
\begin{equation}
U^{\textrm{MS}}=
\begin{pmatrix}
1/\sqrt{2}& 0& 0&i/\sqrt{2} \\
0& 1/\sqrt{2}& i/\sqrt{2}&0 \\
0& i/\sqrt{2}&1/\sqrt{2} &0 \\
i/\sqrt{2}& 0& 0& 1/\sqrt{2}
\end{pmatrix},
\end{equation}
This is a maximally entangling operation. When applied to computational basis states, it generates Bell states; for instance, it transforms $\ket{00}$ to $(\ket{00} + i\ket{11})/\sqrt{2}$. Although locally equivalent to the CNOT gate, the MS gate possesses distinct algebraic properties that may offer advantages for specific algorithmic implementations.

Implementing the non-native Mølmer-Sørensen gate on contemporary superconducting processors requires a compilation strategy that maps its abstract unitary onto the physical device's native gate set. 
We compiled the Mølmer-Sørensen gate into the basis $\{R_Z(\theta), \sqrt{X}, \text{CNOT}\}$, 
resulting in the circuit shown in Fig. \ref{fig:CIRC_MS}. This decomposition was designed to achieve mathematical equivalence to the target unitary while minimizing circuit depth—a critical factor for mitigating error accumulation—resulting in an implementation requiring only a single CNOT gate. Specific rotation angles were calculated to precisely enact the Mølmer-Sørensen transformation while respecting the processor's qubit connectivity constraints.

To characterize the performance of this implementation rigorously, we employed a dual-methodology approach. By executing both direct state measurements (to quantify success probability for specific inputs) and full QPT experiments~\citep{qpt,chuang97,qpt-18,SQSCZ2}. These protocols were run on both a statevector simulator and physical quantum hardware, allowing us to isolate errors arising from hardware noise and decoherence from those inherent to the compilation itself. This multi-faceted  approach provides a complete quantification of the implementation fidelity, capturing both state-specific performance and overall gate quality on current superconducting processors.


\subsection{Direct State Measurements}
\label{sec:StateMeasurements}

To assess the logical performance of the proposed Mølmer-Sørensen gate~\citep{MS-1}, we conducted direct state measurements for a specific computational basis state as input. The gate was applied to the input state \( |00\rangle \),  for which the ideal output is the Bell state \( |\psi_{\text{ideal}}\rangle = \frac{1}{\sqrt{2}}(|00\rangle + i|11\rangle) \).

We first established a theoretical baseline by executing the circuit on a statevector simulator (qasm\_simulator), which provides the expected outcome in the absence of hardware noise. The circuit was then executed on physical hardware using 13,000 measurement shots to ensure statistical significance.

The performance was quantified using the subspace success probability, defined as \(P_{\text{succ}} = p_{|00\rangle} + p_{|11\rangle}\). This metric measures the probability that the output remains within the correct two-dimensional Bell state subspace, providing an operational measure of the gate's functionality and its susceptibility to errors such as state leakage and coherent miscalibration on real hardware~\citep{SPAMerrors}.

\subsection{Complete Gate Characterization via Quantum Process Tomography}
\label{sec:Complete-Characterization}

To obtain a complete characterization of the implemented quantum process, we employed standard QPT. This protocol enables the reconstruction of the full process matrix and a direct fidelity comparison with the ideal gate operation and providing a comprehensive benchmark of gate quality~\citep{qpt,chuang97,qpt-18,SQSCZ2}.

The QPT procedure involved preparing the complete set of linearly independent input states, formed from the tensor product $\{|0\rangle, |1\rangle, |+\rangle, |+i\rangle\}^{\otimes 2}$,  
where \(|+\rangle = \frac{1}{\sqrt{2}}(|0\rangle + |1\rangle)\) and \(|+i\rangle = \frac{1}{\sqrt{2}}(|0\rangle + i|1\rangle)\).
For each input state, we performed quantum state tomography through projective measurements in all two-qubit Pauli bases $\{I, X, Y, Z\}^{\otimes 2}$. Each measurement configuration was executed with 4,000 shots to ensure statistical significance. The same procedure was executed in parallel on a  statevector simulator (qasm\_simulator) to establish a theoretical baseline and on the ibm\_nairobi superconducting processor to assess performance under realistic noise conditions.

The experimental data were used to reconstruct the process matrix $\chi_{\text{exp}}$ which enforces physical constraints to ensure the reconstructed process is completely positive and trace-preserving. The primary performance metric is the process fidelity~\citep{Nielsen02, Horodecki}. This metric provides a rigorous, comprehensive benchmark of gate implementation quality, capturing the combined effect of all error sources during execution.

\begin{figure*}
    \centering
    \includegraphics[width=0.8\textwidth]{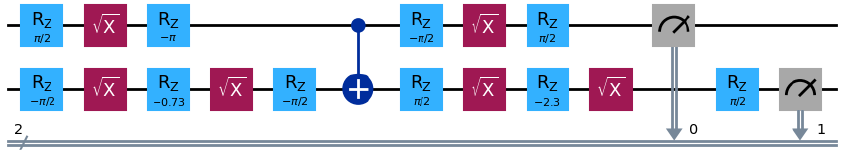}
    \caption{ 
    Hardware-efficient circuit compilation of the Mølmer-Sørensen gate. Decomposition of the Mølmer-Sørensen gate unitary into the native gate set (R$_Z$, $\sqrt{X}$, CNOT) of a superconducting quantum processor. This optimized implementation, requiring only one CNOT gate, enables high-fidelity execution on fixed-architecture hardware. The circuit respects the physical connectivity constraints of the target device.}
    \label{fig:CIRC_MS}
\end{figure*}

\begin{figure}
    \centering
     \includegraphics[width=0.5\textwidth]{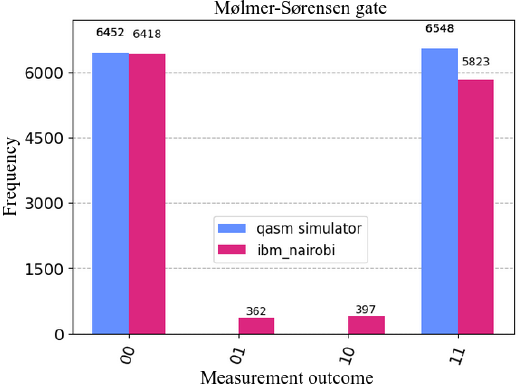}
    \caption{Measurement-based validation of the Mølmer-Sørensen gate on a superconducting quantum processor. Measurement outcomes for the input state $|00\rangle$, comparing results from a noiseless quantum simulator (blue) and hardware execution on ibm\_nairobi (red). The ideal simulated output shows the expected Bell state profile, with population confined to the $|00\rangle$ and $|11\rangle$ states. The hardware results demonstrate a 94.2\% success probability within the correct subspace, with a 5.8\% population leakage into erroneous computational basis states ($|01\rangle$ and $|10\rangle$), quantifying the gate's implementation infidelity. Data from 13,000 shots per execution.}
    \label{fig:measurement_results}
\end{figure}

\begin{figure*}[htbp]
    \centering
    \includegraphics[width=0.8\textwidth]{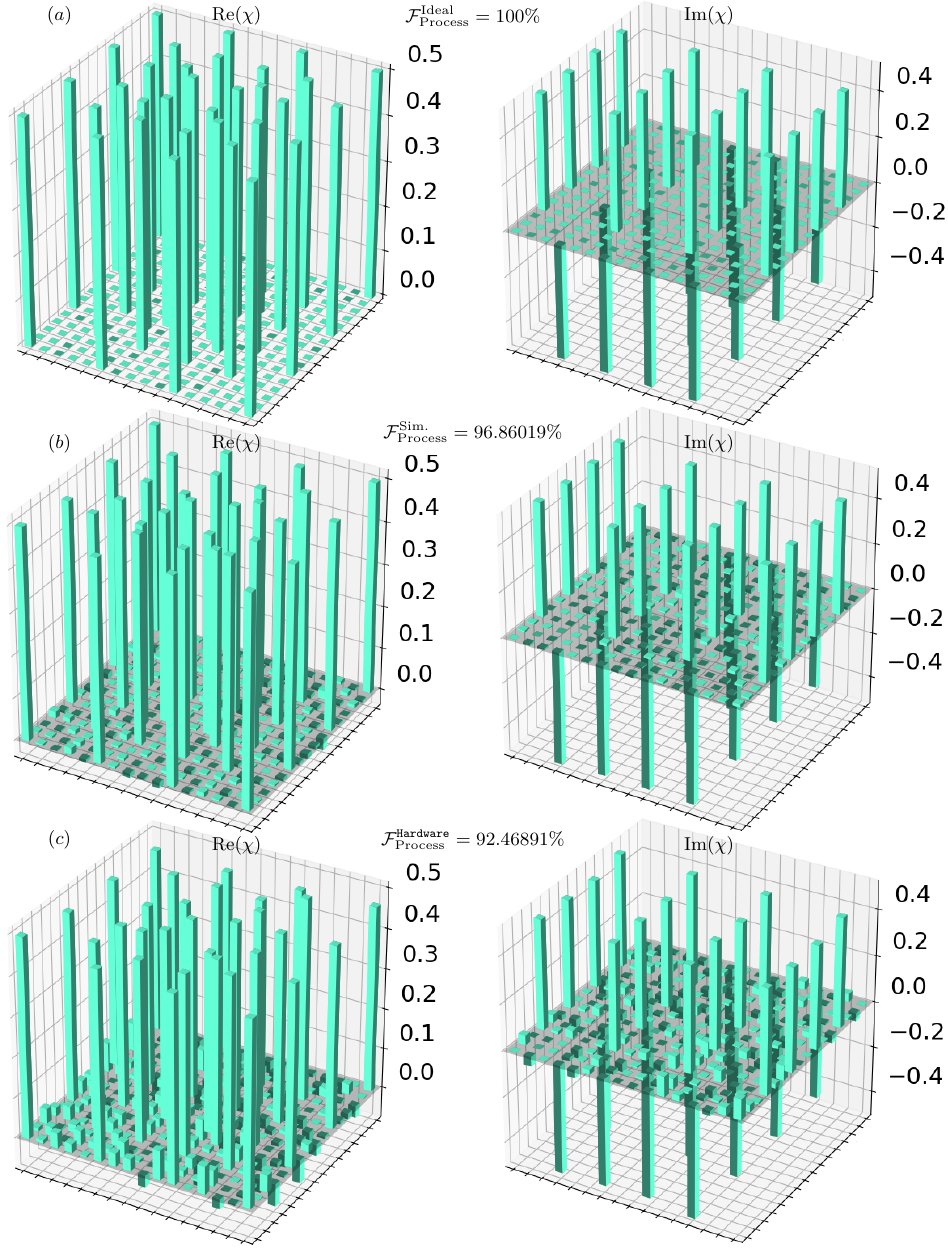}
    \caption{Quantum process tomography of the Mølmer-Sørensen gate. Reconstructed process matrices (Choi matrices) from 
    (a) Ideal theoretical process matrix (\(\mathcal{F}_p^{CZ} = 1.0\)).
    (b) noiseless simulation (process fidelity = 0.969), and (c) hardware execution on ibm\_nairobi (process fidelity = 0.925). The high overlap between the experimental and ideal processes confirms the successful implementation of the target unitary on quantum hardware.}
    \label{fig:MS_tomography}
\end{figure*}

\begin{figure*}
    \centering
    \includegraphics[width=\textwidth]{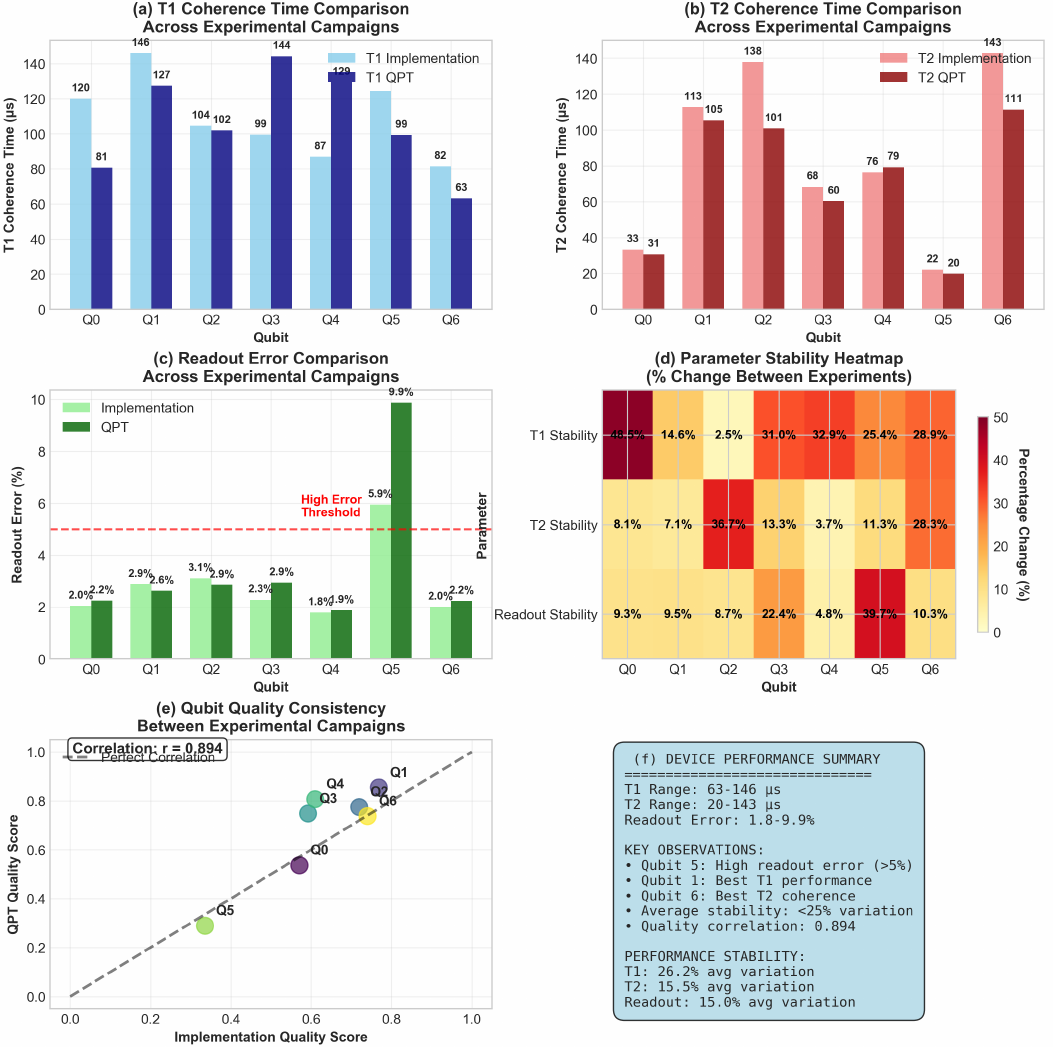}
    \caption{Comprehensive stability analysis of superconducting quantum processor performance across experimental campaigns. 
    (a) T1 coherence time comparison showing 26.2\% average variation between implementation and QPT experiments. (b) T2 dephasing time comparison demonstrating 15.5\% average stability with Q6 exhibiting superior coherence. (c) Readout error analysis revealing consistent performance below 5\% threshold except for Q5. 
    (d) Parameter stability heatmap quantifying percentage variations across all qubits and metrics. 
    (e) Quality correlation scatter plot showing strong consistency between experimental campaigns ($r = 0.894$). (f) Statistical summary highlighting key performance metrics and stability observations. The analysis demonstrates robust device performance suitable for high-fidelity quantum gate implementations, with quality rankings remaining stable despite absolute parameter variations.}
    \label{fig:stability-analysis}
\end{figure*}

\begin{table*}[!htp]
\caption{Quantum processor characterization during Mølmer-Sørensen gate implementation. Device parameters show the hardware conditions for the direct state measurement experiments, highlighting the heterogeneous performance characteristics across qubits. The variation in T2 times ($22–143 \mu$s) and readout errors ($1.8–6.0\%$) across the device establishes the noise environment for the reported gate fidelities and demonstrates consistent gate performance despite qubit-to-qubit parameter variations.}
    \label{tab:qubit_properties_implementation}
    \begin{tabular}{c}
    \includegraphics[width=0.87\textwidth]{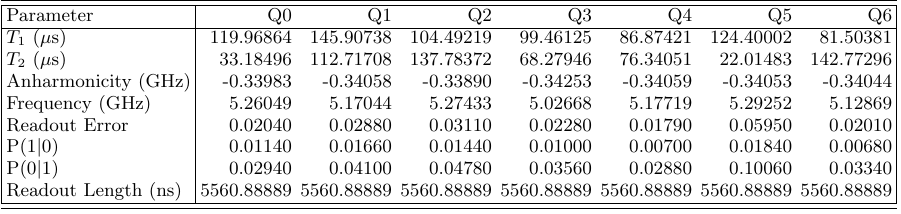} \\
    \end{tabular}
\end{table*}

\begin{table}[!htp]
\caption{Performance comparison of native and compiled two-qubit gates. Process fidelities for the native CX gate and the compiled Mølmer-Sørensen gate, measured via quantum process tomography on a quantum simulator and on real superconducting quantum processor. The Mølmer-Sørensen gate achieves a hardware fidelity (92.47\%) competitive with the native CX gate (93.02\%), demonstrating the efficacy of the hardware-efficient compilation strategy.}
\label{table:FID_comprehensive}
    \begin{tabular}{c}
    \includegraphics[width=0.48\textwidth]{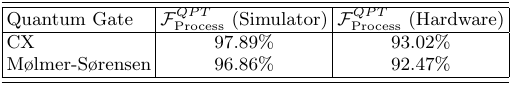} \\
    \end{tabular}
\end{table}

\begin{table*}[!htp]
\caption{Quantum processor characterization during Mølmer-Sørensen gate experiments. Device parameters for the 7-qubit \texttt{ibm\_nairobi} processor, including coherence times ($T_1$, $T_2$), operating frequencies, anharmonicity, and state preparation and measurement (SPAM) errors. This data provides the physical context for the reported gate performance.}
    \label{tab:qubit_properties_QPT}
    \begin{tabular}{c}
    \includegraphics[width=0.87\textwidth]{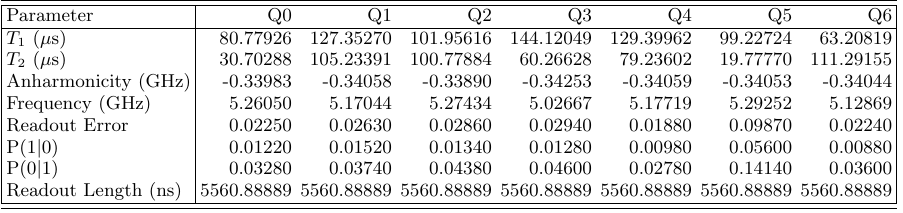} \\
    \end{tabular}
\end{table*}

\section{Results}
\label{sec:Results}

\subsection{Hardware Implementation Efficiency}
\label{sec:logical-efficiency}

We developed a hardware-efficient implementation of the Mølmer-Sørensen gate, compiled into the native gate set of superconducting processors. The resulting circuit, shown in Fig. \ref{fig:CIRC_MS}, consists of strategic single-qubit rotations and a single CNOT gate, optimized to respect processor connectivity constraints. This efficient decomposition allows the non-native MS gate to be executed using the same fundamental operations as standard gates, while providing distinct entangling dynamics.

The experimental outcomes for the Mølmer-Sørensen gate provide a quantitative measure of implementation success, critical for predicting algorithmic performance. For direct state measurements, the probability of obtaining a correct computational basis state after gate execution defines the empirical success probability, \(P_{\text{success}}\).  Applying the gate to the input state \( |00\rangle \), the ideal output is a Bell state, \( |\psi_{\text{ideal}}\rangle = \frac{1}{\sqrt{2}}(|00\rangle + |11\rangle) \), provides a clear metric for logical correctness. 
Execution on a quantum simulator confirmed the theoretical baseline, yielding \(P_{\text{success}}^{\text{(sim)}} \approx 1.0\) for the correct subspace. On hardware, this probability is calculated from the observed populations: \(P_{\text{success}}^{\text{(hw)}} = (P_{00} + P_{11})\), where \(P_{ii}\) are the measured probabilities for each basis state. 

For our implementation, the measured probabilities were \(P_{00} = 0.494\) and \(P_{11} = 0.448\), resulting in a high empirical success probability of \(P_{\text{success}}^{\text{(hw)}} = 0.942\). This corresponds to a state preparation infidelity of \(\epsilon = 1 - P_{\text{success}}^{\text{(hw)}} = 0.058\). The error profile, visualized in Fig.~\ref{fig:measurement_results}, shows this 5.8\% infidelity stems primarily from population leakage into erroneous computational states. This single-gate error rate has direct implications for algorithmic scaling, as the cumulative success probability for a circuit of \(n\) gates would scale as \((1 - \epsilon)^n\), illustrating the exponential decay in output fidelity that constrains the depth of feasible quantum algorithms on NISQ devices. The hardware charechtersistics during our implementaion (Table~\ref{tab:qubit_properties_implementation}) show typical coherence times and error rates for NISQ superconducting processors.

\subsection{Experimental Fidelity Benchmarking}
\label{sec:fidelity}

QPT provides a comprehensive benchmark of our Mølmer-Sørensen gate implementation, revealing performance competitive with native operations on NISQ hardware. The reconstructed process matrices (Fig. \ref{fig:MS_tomography}) visually demonstrate the gate's behavior across different environments, with the experimental matrix maintaining the core structure of the ideal operation despite observable noise effects.

Execution on a statevector simulator yielded a process fidelity of $\mathcal{F}_{\text{proc}}^{\text{sim}} = 96.86\%$. This near-unity value,
\[
\mathcal{F}_{\text{proc}}^{\text{sim}} = 0.9686,
\]
validates the mathematical correctness of our compilation strategy, confirming that the decomposed circuit  accurately implements the target unitary, $U^{\textrm{MS}}$. The minor deviation from unity is attributable to finite-sampling statistics inherent in the tomographic reconstruction with a finite number of measurement shots.

On physical hardware, the Mølmer-Sørensen gate achieves a process fidelity of $\mathcal{F}_{\text{proc}}^{\text{hw}} = 92.47\%$ on ibm\_nairobi. As summarized in Table~\ref{table:FID_comprehensive}, this performance is directly comparable to the 93.02\% fidelity of the device's native CX gate~\citep{CX-CZ}. The fidelity reduction,
\[
\Delta\mathcal{F}_{\text{proc}} = \mathcal{F}_{\text{proc}}^{\text{sim}} - \mathcal{F}_{\text{proc}}^{\text{hw}} \approx 0.044,
\]
quantifies the cumulative effect of device-specific noise processes. The fidelity reduction of approximately 4.4 percentage points between simulation and hardware represents the cumulative effect of device-specific noise processes, including decoherence and control errors. Despite this reduction, the experimental fidelity remains remarkably high, demonstrating that our hardware-efficient compilation strategy—characterized by minimal circuit depth and a single CNOT gate—effectively mitigates the error accumulation that typically plagues complex gate gate decompositions.

The operational robustness of our implementation is further evidenced by its consistent performance across the varied qubit conditions on ibm\_nairobi. As detailed in Table \ref{tab:qubit_properties_QPT}, the device exhibited a range of coherence times ($T_1$ ranging from $63-144 \mu$s) and readout errors ($1.8-9.9\%$) during our experiments. The MS gate's competitive fidelity under these non-ideal conditions establishes it as a viable, high-performance entangling primitive for quantum algorithm design on fixed-architecture superconducting processors~\citep{Chow2011}.

\section{Discussion}
\label{sec:implications}

Our results demonstrate that the Mølmer-Sørensen gate, despite its origins in trapped-ion systems, can be efficiently compiled to achieve performance competitive with native  entangling  gates on superconducting processors. The near-equivalent fidelities of the Mølmer-Sørensen (92.47\%) and the native CX (93.02\%) gates on ibm\_nairobi challenge the conventional assumption that non-native operations necessarily incur substantial fidelity penalties. This finding has several critical implications for compiler design and algorithm implementation in the NISQ era~\citep{G-noise-awarecircuitdesign}.

The high performance of our architecture-adapted Mølmer-Sørensen gate implementation stems from its hardware-efficient decomposition, requiring only a single CNOT gate (Fig.~\ref{fig:CIRC_MS}). This structural efficiency minimizes the accumulation of errors during gate execution, making it particularly suitable for current NISQ hardware~\citep{NISQ18}. The 5.8\% population leakage observed in state-specific measurements (Fig.~\ref{fig:measurement_results}) is consistent with the process fidelity results and aligns with expected error budgets for NISQ superconducting devices, demonstrating methodological consistency across our characterization techniques.

Notably, the Mølmer-Sørensen gate's performance remained robust despite variations in qubit coherence times and readout errors across the processor (Tables~\ref{tab:qubit_properties_implementation} and \ref{tab:qubit_properties_QPT}). This operational stability suggests that the gate's compilation strategy effectively mitigates device-specific noise characteristics, making it a reliable entangling primitive for algorithm design. The comparable performance across different superconducting devices further reinforces its portability across hardware generations.

From a practical perspective, the availability of high-fidelity alternative entangling gates provides quantum algorithm designers with increased flexibility in circuit compilation. Different gates may offer advantages for specific algorithmic primitives or error mitigation strategies. The Mølmer-Sørensen gate's unique entanglement structure may provide more natural or efficient implementations for certain quantum simulations or error correction schemes, or other protocols where its specific interaction dynamics 
are advantageous.

Ultimately, the demonstrated performance parity between native and carefully compiled non-native gates suggests that future quantum compiler optimizations should consider expanded gate sets beyond hardware-native operations. By treating a broader class of efficiently compilable unitaries as viable primitives, we can enable more optimal circuit decompositions and algorithm-specific optimizations. This approach could enable more efficient circuit decompositions and algorithm-specific optimizations in NISQ-era quantum computing~\citep{NISQ18}.

\section{Quantum Hardware Performnace}
\label{sec:Analysis-Hardware}

The comprehensive characterization of the superconducting quantum processor reveals several insights into device performance and stability across experimental campaigns (Figure~\ref{fig:stability-analysis}). The high quality correlation coefficient ($r = 0.894$) between implementation (Table~\ref{tab:qubit_properties_implementation}) and QPT (Table~\ref{tab:qubit_properties_QPT}) campaigns demonstrates remarkable consistency in qubit performance rankings, despite temporal variations in absolute parameter values. This strong correlation indicates that the relative quality hierarchy of qubits remains stable, providing reliable guidance for qubit selection in quantum algorithm design.

The parameter stability analysis reveals distinct patterns across different physical properties. T1 coherence times exhibit the highest variability (26.2\% average variation), reflecting the sensitivity of energy relaxation to environmental fluctuations and calibration conditions. In contrast, T2 dephasing times and readout errors show significantly better stability (15.5\% and 15.0\% variation respectively), suggesting these parameters are more robust to experimental conditions. The exceptional stability of Q2 (2.5\% T1 variation) positions it as a reliable workhorse for quantum operations, while Q0's substantial T1 variability (48.5\%) highlights its susceptibility to environmental noise.

Notably, Qubit 5 maintains consistently low readout errors ($1.8-1.9\%$) despite its challenging coherence properties, demonstrating that readout fidelity can be optimized independently of coherence times. The parameter stability heatmap reveals that while individual qubits may exhibit significant parameter fluctuations, the overall device architecture maintains functional consistency suitable for high-fidelity gate operations.

These findings have direct implications for NISQ-era quantum computing~\citep{NISQ18}. The demonstrated parameter stability supports the feasibility of reproducible quantum experiments, while the identified variations underscore the necessity of dynamic calibration strategies and robust error mitigation techniques. The consistent performance of our compiled Mølmer-Sørensen gate across this variable environment underscores its operational robustness and the effectiveness of our hardware-efficient compilation strategy.

\section{Conclusion}
\label{sec:conclusion}

In this work, we have developed and benchmarked a hardware-efficient implementation of the Mølmer-Sørensen gate for superconducting quantum processors. Our results demonstrate that this non-native entangling gate can achieve performance competitive with hardware-native operations, with a process fidelity of 92.47\% on the ibm\_nairobi processor compared to 93.02\% for the native CX gate. This near-parity in performance establishes that through careful, hardware-aware compilation, the fidelity penalty for employing non-native gates can be minimized to negligible levels on current NISQ devices.

Our study makes two primary contributions. First, we have successfully expanded the practical gate set for superconducting architectures by demonstrating that high-performance gates from other quantum architectures (the trapped-ion domain) can be efficiently compiled without substantial fidelity loss. This provides algorithm designers with a new, effective entangling primitive. Second, we have established a comprehensive benchmarking for cross-platform gate evaluation, offering critical insights into the relationship between compilation strategy, circuit complexity, and realized performance on noisy hardware.

The ability to implement high-fidelity, non-native gates opens new pathways for quantum compiler optimization and algorithm design. As the field advances, the strategic use of a diverse palette of entangling operations---each potentially offering advantages for specific algorithmic primitives or circuit decompositions---will be crucial for maximizing the capabilities of quantum processors. 
Our results suggest that future compiler designs should move beyond a fixed native gate set, instead treating any efficiently compilable unitary as a potential primitive. This approach of an expanded effective gate set promises to be a key enabler for optimizing circuit depth and exploiting hardware-specific advantages on the path toward fault-tolerant quantum computation.

\section*{Acknowledgments}

``We acknowledge the use of IBM Quantum resources in the course of this research. The views and conclusions presented herein are solely those of the author and do not represent the official policies or positions of IBM Quantum or its affiliates."

\section*{Declarations}

\subsection*{Ethical Approval and Consent to participate}
``Not applicable."

\subsection*{Consent for publication}
``All authors have approved the publication. This research did not involve any human, animal or other participants."

\subsection*{Availability of supporting data}
``The datasets generated during and/or analyzed during this study are included within this article."

\subsection*{Competing interests}
``The author declares no competing interests."

\subsection*{Funding}

``No funding, grants, or other financial support were received in connection with this research."

\subsection*{Authors' contributions}

``M. AbuGhanem: Conceptualization, methodology, formal analysis, visualization, Investigation, Validation, Writing, Reviewing and Editing. The author approved the final manuscript."

\end{document}